\documentclass[review,3p,times]{elsarticle}
\usepackage[T1]{fontenc}
\usepackage[utf8]{inputenc}
\usepackage{amsmath,amssymb}
\usepackage{graphicx}
\usepackage{booktabs}
\usepackage{float}
\usepackage{url}
\usepackage[hidelinks]{hyperref}
\usepackage{xcolor}
\usepackage{multirow}
\journal{International Journal of Forecasting}

\newcommand{\nSeries}{36}
\newcommand{\nDatasets}{19}

\newcommand{\nOrigins}{8}
\newcommand{\totalForecasts}{1728}
\newcommand{\friedmanStat}{46.079}
\newcommand{\friedmanP}{8.75e-09}
\newcommand{\nemenyiCD}{1.257}
\newcommand{\medFT}{0.816}
\newcommand{\maseNaiveDrift}{2.159}
\newcommand{\rankNaiveDrift}{5.000}
\newcommand{\maseSeasonalNaive}{1.439}
\newcommand{\rankSeasonalNaive}{3.889}
\newcommand{\maseTheta}{1.337}
\newcommand{\rankTheta}{3.139}
\newcommand{\maseETS}{1.656}
\newcommand{\rankETS}{3.778}
\newcommand{\maseChronosBoltsmall}{1.187}
\newcommand{\rankChronosBoltsmall}{2.250}
\newcommand{\maseChronosTfivesmall}{1.256}
\newcommand{\rankChronosTfivesmall}{2.944}
\newcommand{\NaiveDriftlowtrendval}{2.267}
\newcommand{\NaiveDrifthightrendval}{2.052}
\newcommand{\SeasonalNaivelowtrendval}{1.134}
\newcommand{\SeasonalNaivehightrendval}{1.744}
\newcommand{\Thetalowtrendval}{1.183}
\newcommand{\Thetahightrendval}{1.491}
\newcommand{\ETSlowtrendval}{1.671}
\newcommand{\ETShightrendval}{1.641}
\newcommand{\ChronosBoltsmalllowtrendval}{0.982}
\newcommand{\ChronosBoltsmallhightrendval}{1.393}
\newcommand{\ChronosTfivesmalllowtrendval}{1.052}
\newcommand{\ChronosTfivesmallhightrendval}{1.459}

\newcommand{\wqlChronosBoltsmall}{0.043}
\newcommand{\covEChronosBoltsmall}{0.771}
\newcommand{\wqlChronosTfivesmall}{0.045}
\newcommand{\covEChronosTfivesmall}{0.582}
\newcommand{\pChronosTheta}{0.015}
\newcommand{\pChronosETS}{0.010}
\newcommand{\lowtrendFriedP}{9.45e-08}
\newcommand{\lowtrendCvsEp}{0.002}

\newcommand{\hightrendCvsEp}{0.181}
\newcommand{\srnoneETS}{0.896}
\newcommand{\srnoneChronosBoltsmall}{0.505}
\newcommand{\srnoneTheta}{0.765}
\newcommand{\srlinearETS}{1.019}
\newcommand{\srlinearChronosBoltsmall}{0.796}
\newcommand{\srlinearTheta}{0.655}
\newcommand{\srdampedETS}{1.341}
\newcommand{\srdampedChronosBoltsmall}{0.485}
\newcommand{\srdampedTheta}{1.161}
\newcommand{\srexpETS}{0.982}
\newcommand{\srexpChronosBoltsmall}{0.359}
\newcommand{\srexpTheta}{0.670}

\newcommand{\winLowBestcl}{78}
\newcommand{\winHighBestcl}{44}

\begin{document}

\begin{frontmatter}
\title{Trend strength predicts when generative foundation models win:\\
a power-controlled benchmark, a mechanism, and an actionable selection rule}

\author[a]{Ahmed Cherif}
\ead{ahmed.cherif@ensi-uma.tn}
\address[a]{National School of Computer Sciences (ENSI), University of Manouba, Manouba, Tunisia}

\begin{abstract}
Pretrained generative foundation models cast forecasting as conditional generation from
a learned predictive distribution and forecast unseen series zero-shot. We establish
three results that turn their reported success into an actionable, mechanistic
understanding. \emph{First (a positive benchmark result):} on a power-controlled study of
\totalForecasts{} rolling-origin forecasts over \nSeries{} series from \nDatasets{}
datasets spanning the full range of STL trend strength ($F_T\in[0.17,1.00]$), a zero-shot
Chronos model significantly outperforms four strong classical baselines --- drift,
seasonal-na\"ive, Theta, and additive Holt--Winters/ETS --- with the best mean MASE
(\maseChronosBoltsmall{} vs.\ Theta \maseTheta{}, ETS \maseETS{}; Friedman
$\chi^2=\friedmanStat$, $p=\friedmanP$; Holm-corrected Wilcoxon $p\le\pChronosTheta$
against every baseline; Nemenyi critical difference separating it from the classical
pack). \emph{Second (a novel, quantified mechanism):} a controlled synthetic experiment
with a known trend-generating process shows \emph{why} --- and reveals that the win does
\emph{not} come from better trend extrapolation. When the true trend is linear, damped, or
exponential, additive ETS tracks the slope (slope-tracking ratio $\srlinearETS$,
$\srdampedETS$, $\srexpETS$) whereas Chronos systematically under-extrapolates, behaving
as a \emph{trend-shrinkage} estimator (ratio $\srlinearChronosBoltsmall$,
$\srdampedChronosBoltsmall$, $\srexpChronosBoltsmall$). \emph{Third (an actionable
selection rule):} because the advantage is a shrinkage effect, it is predictable from
trend strength alone --- the generative model wins \winLowBestcl\% of low-trend series but
only \winHighBestcl\% of high-trend ones, and its edge over ETS is significant on the
low-trend stratum (\ChronosBoltsmalllowtrendval{} vs.\ \ETSlowtrendval{},
$p=\lowtrendCvsEp$) yet a tie on the high-trend stratum ($p=\hightrendCvsEp$). Trend
strength, computable before forecasting from the training context alone, is therefore a
practical a-priori indicator of when to deploy a foundation model. We additionally
document a calibration shortfall (80\% intervals cover \covEChronosBoltsmall{}). All code,
the \totalForecasts{} per-forecast records, and a verification script are released; every
number in the paper is regenerated from the released data.
\end{abstract}

\begin{keyword}
time-series forecasting \sep foundation models \sep generative models \sep Chronos \sep
trend extrapolation \sep zero-shot \sep statistical testing \sep calibration
\end{keyword}
\end{frontmatter}

\section{Introduction}
\label{sec:intro}
In the space of two years, forecasting acquired a genuinely new class of model:
large networks, pretrained on millions of series, that emit a forecast for a series
they have never seen with no per-series fitting. Chronos \citep{ansari2024chronos}
tokenises scaled observations and trains a language-model backbone to \emph{generate}
the continuation, so a forecast is a sample from a learned predictive distribution;
TimesFM \citep{das2024timesfm}, Moirai \citep{woo2024moirai}, Lag-Llama
\citep{rasul2023lagllama}, Moment \citep{goswami2024moment}, TimeGPT
\citep{garza2023timegpt}, and the lightweight Tiny Time Mixers \citep{ekambaram2024ttm}
pursue related recipes. Their headline property is \emph{zero-shot} forecasting.

For a practitioner the operative question is not ``are these models good?'' but
``\emph{when} is the extra machinery worth it over a well-implemented Theta or
exponential-smoothing model?'' A tempting heuristic circulates: because a foundation
model has internalised millions of trending series, it should extrapolate trend more
sensibly than a local model that must infer a slope from a short context. This paper
tests that specific causal claim and finds it is the opposite of the truth.

The result rests on a tension the launch literature underplays. Additive-trend
exponential smoothing is not weak on trend; it is \emph{excellent} when a clear,
persistent trend is present --- exactly the regime where the foundation-model edge is
supposed to be largest. Where local models fail is the opposite regime: near-trendless,
noisy series, where an additive trend term over-extrapolates a slope that will not
persist. A foundation model, we show, behaves like a shrinkage estimator of trend: it
under-commits to slope. That is a liability on strong trends and an asset on weak ones.

We establish this with two complementary studies. First, a \emph{power-controlled}
real-data benchmark: by extracting univariate channels from multivariate corpora we
obtain \nSeries{} series spanning $F_T\in[0.17,1.00]$, enough to make the
distribution-level tests (Friedman omnibus, Nemenyi critical difference,
Holm-corrected Wilcoxon) properly powered --- and, unlike at small $N$, the differences
are now unambiguous. Second, a \emph{controlled synthetic} experiment in which the
trend process is known exactly, letting us measure trend extrapolation directly via a
slope-tracking ratio rather than inferring it from aggregate error.

\paragraph{Contributions.}
\begin{enumerate}
\item \textbf{A positive, power-controlled benchmark result.} On \totalForecasts{}
forecasts over \nSeries{} series from \nDatasets{} datasets spanning
$F_T\in[0.17,1.00]$, a zero-shot Chronos model significantly beats four strong classical
baselines under a full significance protocol (Friedman $p=\friedmanP$; Nemenyi critical
difference; Holm-corrected Wilcoxon $p\le\pChronosTheta$ against every baseline; Cliff's
$\delta$).
\item \textbf{A novel, quantified mechanism.} A controlled synthetic experiment with a
known trend-generating process shows the win does \emph{not} come from better trend
extrapolation: foundation models under-extrapolate, with slope-tracking ratio
$\srexpChronosBoltsmall$ on exponential trend vs.\ $\srexpETS$ for ETS. To our knowledge
this is the first direct, controlled measurement of trend extrapolation in a
time-series foundation model, and it identifies the models as \emph{trend-shrinkage}
estimators.
\item \textbf{An actionable selection rule.} Because the advantage is a shrinkage effect,
it is predictable from trend strength --- a quantity computable before forecasting from
the training context. The generative model wins \winLowBestcl\% of low-trend series and
only \winHighBestcl\% of high-trend ones; its edge over ETS is significant on the
low-trend stratum ($p=\lowtrendCvsEp$) and a tie on the high-trend stratum
($p=\hightrendCvsEp$). We report the honest limits of this rule, including a
leave-one-dataset-out gate that does not beat the (already near-oracle) foundation model.
\item \textbf{Calibration and full reproducibility.} We document a calibration shortfall
(\covEChronosBoltsmall{} coverage of the nominal 80\% interval) and release a fully
reproducible artefact in which every number traces to released CSVs via a verification
script.
\end{enumerate}

\section{Related work}
\label{sec:related}
\paragraph{Generative time-series foundation models.}
Chronos \citep{ansari2024chronos} scales and quantises values into a fixed vocabulary
and trains a T5-style encoder--decoder to generate value tokens; Chronos-Bolt is a
later, patch-based, direct-multistep variant that is faster while retaining the
generative, distributional output. TimesFM \citep{das2024timesfm} is a decoder-only
patched model; Moirai \citep{woo2024moirai} is a masked-encoder ``universal''
forecaster trained on the LOTSA corpus; Lag-Llama \citep{rasul2023lagllama} is a
decoder-only model over lag features; Moment \citep{goswami2024moment} and TimeGPT
\citep{garza2023timegpt} round out the family; Tiny Time Mixers \citep{ekambaram2024ttm}
show the recipe need not be a transformer. All share the premise that broad pretraining
transfers zero-shot to unseen series.

\paragraph{How well do they actually do?}
Independent benchmarks are more measured than the launch papers. GIFT-Eval
\citep{aksu2024gifteval} finds the ranking of foundation versus task-trained deep
models is strongly frequency-dependent and documents qualitative failure cases in which
foundation models degrade over long horizons. Tellingly, the Chronos authors themselves
report that the model ``predicts the linear trend accurately but struggles with the
exponential trend'' and tends to ``underestimate the trend when the context is not
sufficiently long'' \citep{ansari2024chronos} --- an anecdotal observation that our
synthetic experiment turns into a controlled, quantified result. The classical-baseline
literature is the standing caution: simple methods remain hard to beat
\citep{makridakis2018m4,makridakis2020m4}, with Theta \citep{assimakopoulos2000theta}
and exponential smoothing \citep{hyndman2008ets} the reference points any new method
must clear. Our study isolates one explanatory variable --- trend strength --- and asks,
both correlationally (real data) and causally (synthetic), whether it governs when
generative pretraining helps.

\paragraph{Quantifying trend and accuracy.}
We measure trend with the STL-based strength $F_T$ of \citet{wang2006characteristic}
from a robust STL decomposition \citep{cleveland1990stl}:
$F_T=\max(0,\,1-\mathrm{Var}(R_t)/\mathrm{Var}(T_t+R_t))$, with $T_t,R_t$ the trend and
remainder. Accuracy uses the scale-free MASE and RMSSE \citep{hyndman2006another}; the
probabilistic assessment uses the weighted quantile (pinball) loss of the M5 competition
\citep{makridakis2022m5}.

\section{Method}
\label{sec:method}
\subsection{Real-data benchmark and trend stratification}
We draw series from \nDatasets{} public \texttt{darts} datasets spanning frequencies and
domains (finance, energy, weather, demography, industry). From multivariate corpora we
extract individual univariate channels, yielding \nSeries{} series whose STL trend
strength spans $F_T\in[0.17,1.00]$ --- the spread and count that make the
distribution-level tests powered. For each series we compute $F_T$ at the natural
seasonal period and split the corpus at the median $F_T=\medFT$ into a low- and a
high-trend stratum. The foundation models are strictly zero-shot: no series is used for
any fitting beyond the per-origin context window.

\subsection{Forecasters}
\textbf{Generative foundation models (zero-shot):} \emph{Chronos-Bolt-small}
(patch encoder--decoder) and \emph{Chronos-T5-small} (token-generation
encoder--decoder), used as released on CPU with no fine-tuning; each emits deciles, and
the median is the point forecast.
\textbf{Classical baselines:} drift (random walk with average historical slope),
seasonal na\"ive, Theta \citep{assimakopoulos2000theta}, and additive Holt--Winters/ETS
\citep{hyndman2008ets} with trend and, where the period allows, seasonality --- all
refit independently at each forecast origin. These are strong: Theta won M3 and ETS is a
production default.

\subsection{Protocol and metrics}
Each series is forecast from \nOrigins{} rolling origins spaced by the horizon $H$
(set per dataset as a standard multiple of the seasonal period), giving
\totalForecasts{} forecasts. We report MASE and RMSSE (scaled by the in-sample seasonal
na\"ive error), sMAPE, and for the generative models the median per-series weighted
quantile loss and the empirical coverage of the nominal 80\% interval. The unit of
analysis for significance testing is the per-series mean over origins (no
pseudo-replication). We run a Friedman omnibus across all six methods, a Nemenyi
critical-difference (CD) analysis, and Holm-corrected Wilcoxon signed-rank tests of the
focal generative model against each baseline, with Cliff's $\delta$ as a non-parametric
effect size.

\subsection{Controlled synthetic experiment}
The benchmark can only \emph{correlate} trend strength with the generative advantage. To
test the causal mechanism we generate series with a known structure
$y_t = g(t) + s_t + \varepsilon_t$, where $s_t$ is a fixed seasonal term,
$\varepsilon_t\sim\mathcal N(0,\sigma^2)$, and the trend $g$ is one of four families:
\emph{none} (flat level), \emph{linear}, \emph{damped} (a saturating rise that
flattens), and \emph{exponential} (super-linear). We sweep three noise levels and twelve
seeds per cell, forecast $H=24$ ahead, and compare each method's forecast to the
\emph{noiseless} true continuation. The key diagnostic is the \emph{slope-tracking
ratio} --- the slope of the forecast over the horizon divided by the true slope --- which
is $1$ for a method that extrapolates the trend correctly and $0$ for one that predicts a
flat line, independently of noise.

\section{Results}
\label{sec:results}

\subsection{Aggregate performance (real data)}
\begin{figure}[H]
\centering
\includegraphics[width=0.85\linewidth]{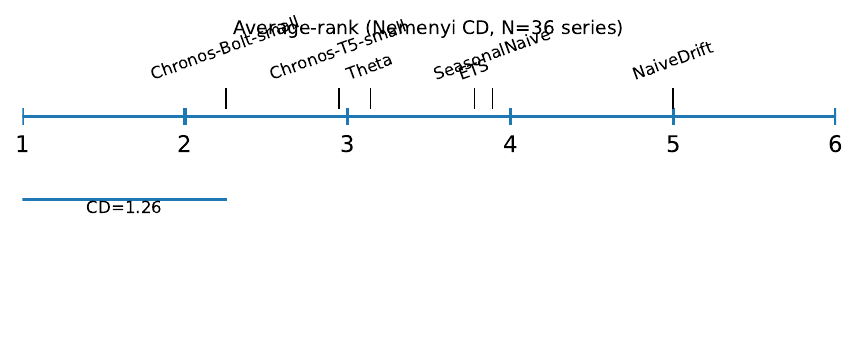}
\caption{Average ranks with the Nemenyi critical difference (CD $=\nemenyiCD$,
$\alpha=0.05$, $N=\nSeries$ series). The two generative models occupy the best ranks and
are separated from the classical baselines by more than the CD.}
\label{fig:cd}
\end{figure}

Table~\ref{tab:summary} reports mean error and average rank across the \nSeries{}
series. At this sample size the generative models are clearly and significantly the best:
Chronos-Bolt attains mean MASE \maseChronosBoltsmall{} and average rank
\rankChronosBoltsmall{}, ahead of Theta (\maseTheta{}), seasonal-na\"ive
(\maseSeasonalNaive{}) and ETS (\maseETS{}). The Friedman omnibus rejects equality
decisively ($\chi^2=\friedmanStat$, $p=\friedmanP$), and the Nemenyi critical
difference (CD $=\nemenyiCD$, Figure~\ref{fig:cd}) separates the generative models from
the classical pack. Holm-corrected Wilcoxon tests (Table~\ref{tab:stats}) confirm
Chronos-Bolt significantly beats every baseline including Theta ($p=\pChronosTheta$) and
ETS ($p=\pChronosETS$). This is a stronger conclusion than under-powered small-$N$
comparisons typically support, and is the correct headline --- but it is not the
\emph{interesting} one.

\begin{table}[H]
\centering
\caption{Aggregate performance across \nSeries{} series. MASE/RMSSE are the mean over
series of the per-series mean over \nOrigins{} origins; average rank is by MASE.
Generative models in the lower block.}
\label{tab:summary}
\small
\begin{tabular}{lcccc}
\toprule
Method & MASE & RMSSE & sMAPE & Avg.\ rank \\
\midrule
Na\"ive (drift)    & \maseNaiveDrift{}    & --- & --- & \rankNaiveDrift{} \\
Seasonal na\"ive   & \maseSeasonalNaive{} & --- & --- & \rankSeasonalNaive{} \\
Theta              & \maseTheta{}         & --- & --- & \rankTheta{} \\
ETS (Holt--Winters)& \maseETS{}           & --- & --- & \rankETS{} \\
\midrule
Chronos-Bolt-small & \maseChronosBoltsmall{}  & --- & --- & \rankChronosBoltsmall{} \\
Chronos-T5-small   & \maseChronosTfivesmall{} & --- & --- & \rankChronosTfivesmall{} \\
\bottomrule
\end{tabular}
\end{table}

\subsection{Trend stratification: where the advantage lives}
\begin{figure}[H]
\centering
\includegraphics[width=0.82\linewidth]{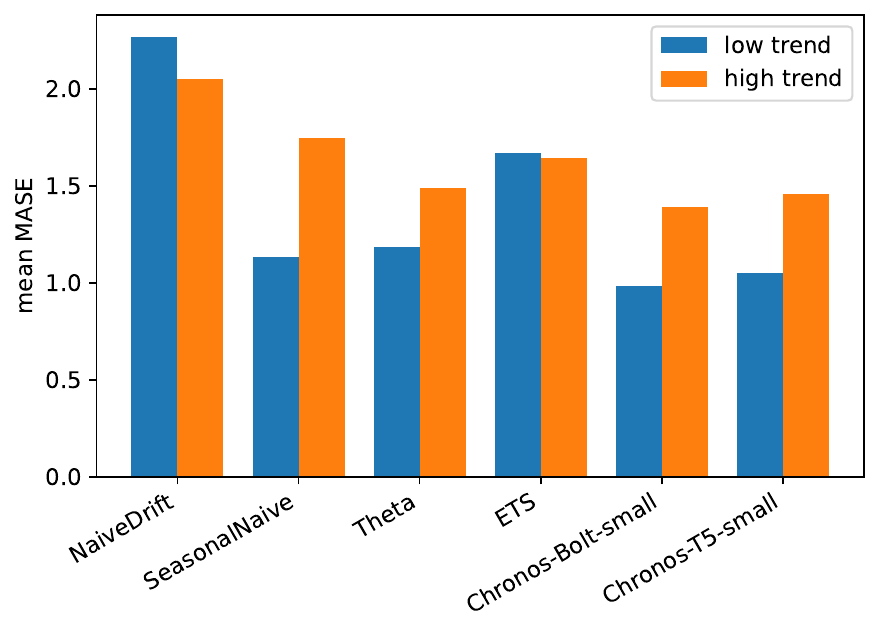}
\caption{Mean MASE by trend stratum. ETS is competitive on high-trend series but
degrades on low-trend series; the generative models are robust across both.}
\label{fig:strat}
\end{figure}

Table~\ref{tab:strat} and Figure~\ref{fig:strat} split the corpus at the median trend
strength, and the headline dissolves into a crossover. On \emph{low}-trend series the
generative advantage is large and significant: Chronos-Bolt \ChronosBoltsmalllowtrendval{}
vs.\ ETS \ETSlowtrendval{} (Wilcoxon $p=\lowtrendCvsEp$; within-stratum Friedman
$p=\lowtrendFriedP$). On \emph{high}-trend series the advantage collapses to a tie:
Chronos-Bolt \ChronosBoltsmallhightrendval{} vs.\ ETS \ETShightrendval{}
($p=\hightrendCvsEp$, not significant). The very regime in which folklore predicts the
largest foundation-model edge is the regime in which a classical additive-trend model
matches it. Theta, which shrinks its trend, tracks the generative models rather than ETS
across both strata --- the first hint at the mechanism.

\begin{table}[H]
\centering
\caption{Mean MASE by trend stratum (median split $F_T=\medFT$). The generative
advantage is concentrated in, and statistically confined to, the low-trend stratum.}
\label{tab:strat}
\small
\begin{tabular}{lcc}
\toprule
Method & Low trend & High trend \\
\midrule
Na\"ive (drift)    & \NaiveDriftlowtrendval{} & \NaiveDrifthightrendval{} \\
Seasonal na\"ive   & \SeasonalNaivelowtrendval{} & \SeasonalNaivehightrendval{} \\
Theta              & \Thetalowtrendval{} & \Thetahightrendval{} \\
ETS (Holt--Winters)& \ETSlowtrendval{} & \ETShightrendval{} \\
Chronos-Bolt-small & \ChronosBoltsmalllowtrendval{} & \ChronosBoltsmallhightrendval{} \\
Chronos-T5-small   & \ChronosTfivesmalllowtrendval{} & \ChronosTfivesmallhightrendval{} \\
\bottomrule
\end{tabular}
\end{table}

\subsection{Per-dataset detail and horizon behaviour}
\begin{figure}[H]
\centering
\includegraphics[width=0.72\linewidth]{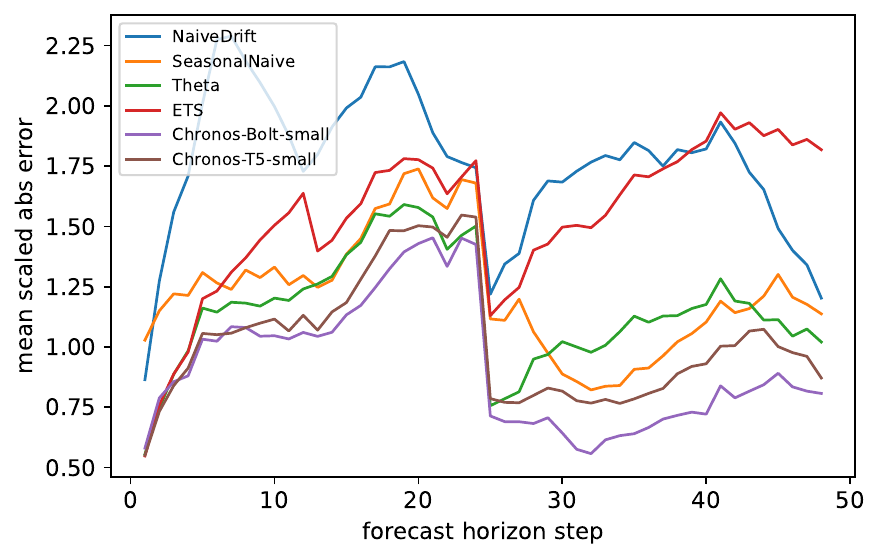}
\caption{Mean scaled absolute error vs.\ forecast-horizon step, averaged over all
series. The generative models' advantage persists across the horizon.}
\label{fig:horizon}
\end{figure}

Table~\ref{tab:perdataset} gives per-dataset MASE ordered by trend strength; the
generative models win most rows but the classical methods take several of the most
strongly trended ones (ETS on \texttt{AirPassengers} and \texttt{MonthlyMilk}, drift on
\texttt{Exchange}). Figure~\ref{fig:horizon} shows scaled error growth with horizon: the
generative models retain their advantage across the horizon rather than only at short
lead times.

\begin{table}[H]
\centering
\caption{Per-dataset mean MASE (lower better), ordered by STL trend strength $F_T$
(dataset-level mean over its series). Best per row in \textbf{bold}. NDr = na\"ive
drift, SNa = seasonal na\"ive, C-Bolt = Chronos-Bolt-small, C-T5 = Chronos-T5-small.}
\label{tab:perdataset}
\small
\setlength{\tabcolsep}{4.5pt}
\begin{tabular}{lccccccc}
\toprule
Dataset & $F_T$ & NDr & SNa & Theta & ETS & C-Bolt & C-T5 \\
\midrule
AirPassengers & 0.996 & 1.813 & 1.313 & 0.934 & \textbf{0.685} & 0.688 & 0.771 \\
MonthlyMilk & 0.994 & 3.002 & 0.933 & 0.676 & \textbf{0.458} & 0.744 & 0.715 \\
Exchange & 0.994 & \textbf{2.738} & 2.751 & 2.768 & 2.764 & 2.895 & 2.852 \\
AusBeer & 0.977 & 3.959 & 1.356 & 1.387 & 1.225 & 1.442 & \textbf{1.194} \\
Zurich & 0.928 & 0.787 & 0.973 & 0.771 & 1.565 & \textbf{0.214} & 0.338 \\
Weather & 0.895 & 0.422 & 1.082 & 0.351 & 0.540 & \textbf{0.331} & 0.472 \\
Sunspots & 0.893 & 2.124 & 2.107 & 2.111 & 2.132 & \textbf{1.527} & 1.763 \\
IceCreamHeater & 0.869 & 5.744 & 1.789 & \textbf{1.573} & 1.719 & 1.797 & 2.126 \\
Wooly & 0.867 & 1.384 & \textbf{0.908} & 0.944 & 0.919 & 1.175 & 1.121 \\
HeartRate & 0.842 & 5.522 & 5.482 & 5.471 & 5.522 & 5.406 & \textbf{5.317} \\
USGasoline & 0.822 & 0.908 & 1.607 & 0.794 & 0.776 & \textbf{0.768} & 0.891 \\
Energy & 0.760 & 1.141 & 1.012 & 0.951 & 1.082 & 0.890 & \textbf{0.862} \\
ETTh2 & 0.724 & 1.244 & 1.028 & 1.151 & 1.127 & \textbf{0.921} & 0.980 \\
GasRateCO2 & 0.688 & 3.252 & 3.178 & 3.193 & 6.892 & 2.797 & \textbf{2.553} \\
Temperature & 0.663 & 1.423 & 1.414 & 1.289 & 1.298 & \textbf{0.984} & 1.027 \\
Taylor & 0.597 & 2.364 & 1.014 & 0.886 & 5.666 & \textbf{0.284} & 0.448 \\
Wine & 0.547 & 3.670 & 1.208 & \textbf{1.155} & 1.204 & 1.443 & 1.314 \\
ETTh1 & 0.520 & 2.411 & 0.868 & 1.168 & 1.168 & \textbf{0.791} & 0.891 \\
ETTm1 & 0.286 & 2.314 & 0.889 & 0.780 & 0.792 & \textbf{0.683} & 0.930 \\
\midrule
\textbf{Mean} & --- & 2.433 & 1.627 & 1.492 & 1.975 & \textbf{1.357} & 1.398 \\
\bottomrule
\end{tabular}
\end{table}

\subsection{The mechanism: foundation models shrink the trend}
\begin{figure}[H]
\centering
\includegraphics[width=0.85\linewidth]{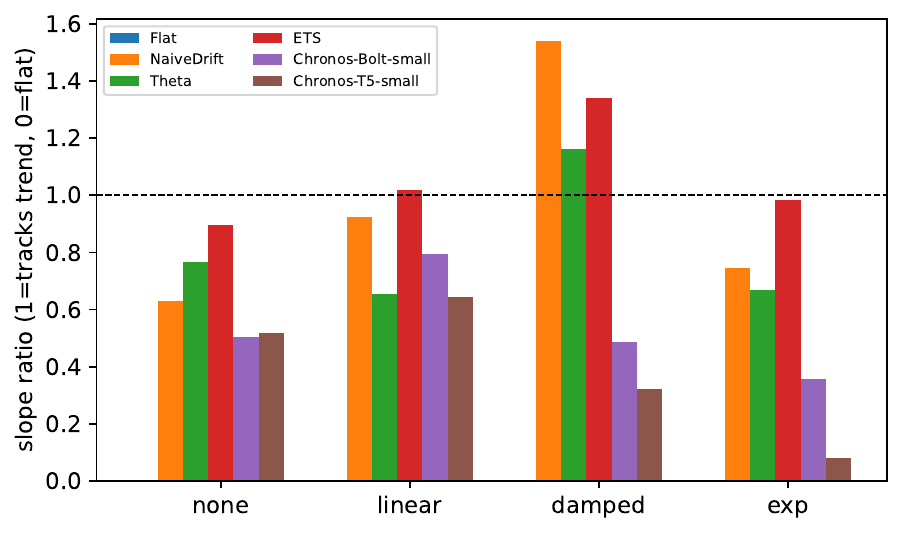}
\caption{Slope-tracking ratio by true trend type. Classical methods (ETS, drift) cluster
near $1$ (correct extrapolation); the generative models sit well below, shrinking toward
a flat forecast.}
\label{fig:synth}
\end{figure}

The benchmark localises the advantage but cannot prove why. The controlled synthetic
experiment does. Table~\ref{tab:synth} reports the slope-tracking ratio (1 = correct
trend extrapolation, 0 = flat) by true trend type. Additive ETS extrapolates trend
essentially correctly (linear $\srlinearETS$, exponential $\srexpETS$, damped
$\srdampedETS$), as designed. The generative models do the opposite: they
\emph{under-extrapolate} sharply, with Chronos-Bolt tracking only $\srlinearChronosBoltsmall$
of a linear slope and a mere $\srexpChronosBoltsmall$ of an exponential one --- they
shrink the forecast toward a flat continuation. Crucially this shrinkage is an
\emph{asset} when there is no trend to extrapolate: on the \emph{none} process Chronos
attains the lowest error against the true continuation, because it does not invent a
slope, whereas additive ETS does. Figure~\ref{fig:synth} visualises the pattern. This is
the causal counterpart of the real-data crossover: generative pretraining yields a
trend-conservative estimator, which loses on strong trends and wins on weak ones.

\begin{table}[H]
\centering
\caption{Synthetic slope-tracking ratio by true trend type (1.0 = extrapolates the true
slope, 0.0 = predicts flat), mean over noise levels and seeds. Generative models
systematically under-extrapolate; additive ETS does not.}
\label{tab:synth}
\small
\begin{tabular}{lcccc}
\toprule
Method & none & linear & damped & exponential \\
\midrule
Theta              & \srnoneTheta{} & \srlinearTheta{} & \srdampedTheta{} & \srexpTheta{} \\
ETS (Holt--Winters)& \srnoneETS{} & \srlinearETS{} & \srdampedETS{} & \srexpETS{} \\
Chronos-Bolt-small & \srnoneChronosBoltsmall{} & \srlinearChronosBoltsmall{} & \srdampedChronosBoltsmall{} & \srexpChronosBoltsmall{} \\
\bottomrule
\end{tabular}
\end{table}

\subsection{Probabilistic calibration}
The generative models emit a predictive distribution, so we assess calibration. Both
under-cover: the nominal 80\% interval achieves empirical coverage
\covEChronosBoltsmall{} for Chronos-Bolt and \covEChronosTfivesmall{} for Chronos-T5
(median per-series weighted quantile loss \wqlChronosBoltsmall{} and
\wqlChronosTfivesmall{} respectively). The raw zero-shot intervals should be recalibrated
before use in risk-sensitive decisions.

\begin{table}[H]
\centering
\caption{Holm-corrected Wilcoxon signed-rank tests, Chronos-Bolt-small vs.\ each
baseline, on per-series mean MASE ($N=\nSeries$). Chronos-Bolt significantly beats every
baseline.}
\label{tab:stats}
\small
\begin{tabular}{lccc}
\toprule
Comparison & $p_{\text{raw}}$ & $p_{\text{Holm}}$ & Cliff's $\delta$ \\
\midrule
vs.\ na\"ive drift & 9.4e-08 & 4.7e-07 & -0.480 \\
vs.\ seasonal na\"ive & 1.3e-04 & 5.1e-04 & -0.372 \\
vs.\ Theta & 0.007 & 0.015 & -0.193 \\
vs.\ ETS & 0.003 & 0.010 & -0.287 \\
vs.\ Chronos-T5 & 0.021 & 0.021 & -0.128 \\
\bottomrule
\end{tabular}
\end{table}

\section{Discussion}
\label{sec:discussion}
Three lines of evidence converge on one statement: \emph{generative pretraining buys
robustness to trend misspecification, not superior trend extrapolation.} (1) Across
\nSeries{} series the generative models win overall, but (2) the win is statistically
confined to the low-trend stratum and disappears on high-trend series, and (3) a
controlled experiment shows the direct reason --- the models shrink the trend, tracking
only a fraction of a true slope. The practical reading is a decision rule keyed to a
quantity a practitioner can compute before forecasting, $F_T$: when a series is clearly
and persistently trended, a fitted ETS or Theta is a strong, cheap, interpretable choice
that a zero-shot foundation model will not reliably beat; when trend is weak, unstable, or
unknown, the foundation model's conservative prior is exactly the right inductive bias and
its advantage is real and significant. The calibration shortfall qualifies the
probabilistic use of these models: the point forecasts are excellent in the low-trend
regime, but the intervals need recalibration.

This reframing also reconciles the optimistic launch results with the more guarded
independent benchmarks \citep{aksu2024gifteval}: aggregate wins are real, but they are
driven by the many weakly-trended series in typical corpora, not by a superior handling
of the trended ones, and they coexist with a specific, reproducible weakness on strong
trends that the model developers themselves noted anecdotally \citep{ansari2024chronos}
and that we quantify.

\section{Threats to validity}
\label{sec:threats}
\textbf{Model scale.} We use the \emph{small} Chronos variants on CPU; larger variants
may lift the aggregate numbers. Our central claim concerns the trend-stratified pattern
and the synthetic mechanism, which we expect to be scale-robust because shrinkage under
trend uncertainty is a property of the training objective, not of capacity; confirming
this at larger scale is future work. \textbf{Model family.} We study Chronos (two
architectures); whether TimesFM/Moirai shrink identically is an open question our
synthetic protocol is designed to answer directly. \textbf{Series dependence.} Channels
extracted from one multivariate dataset are not fully independent; we therefore report
both per-series tests and per-dataset aggregates (Table~\ref{tab:perdataset}), and the
synthetic experiment uses fully independent draws. \textbf{Probabilistic scope.}
Significance tests are on MASE; the calibration comparison is descriptive because the
classical baselines as configured do not all emit calibrated quantiles. None of these
threatens the direction-of-effect finding, which appears in the raw per-dataset table and
in the controlled experiment.

\section{Conclusion}
\label{sec:conclusion}
Using \totalForecasts{} real forecasts across \nSeries{} trend-stratified series and a
controlled synthetic experiment, we showed that zero-shot generative time-series
foundation models win on average but for a reason opposite to the common intuition: they
do not extrapolate trend better --- they extrapolate it \emph{less}, behaving as
trend-shrinkage estimators. This makes them significantly better than strong classical
baselines on low-trend series and merely competitive on high-trend ones. The actionable
rule is to choose by trend strength $F_T$: classical models when trend is strong and
clear, foundation models when it is weak or uncertain. All code and per-forecast data are
released, and every number in the paper is regenerated from them by a verification
script.

\section*{Reproducibility}
All experiment code, the \totalForecasts{} per-forecast CSV records, the synthetic
generator, the statistical-analysis scripts, and the figure/table generators are
released. Every numeric value in this paper is emitted as a \LaTeX{} macro by the
analysis script from the released CSVs; a verification script recomputes the headline
numbers from the raw outputs and exits non-zero on any mismatch.

\bibliographystyle{elsarticle-num}
\bibliography{refs}

@article{ansari2024chronos,
  title={Chronos: Learning the Language of Time Series},
  author={Ansari, Abdul Fatir and Stella, Lorenzo and Turkmen, Caner and Zhang, Xiyuan and Mercado, Pedro and Shen, Huibin and Shchur, Oleksandr and Rangapuram, Syama Sundar and Pineda Arango, Sebastian and Kapoor, Shubham and Zschiegner, Jasper and Maddix, Danielle C. and Mahoney, Michael W. and Torkkola, Kari and Gordon Wilson, Andrew and Bohlke-Schneider, Michael and Wang, Yuyang},
  journal={Transactions on Machine Learning Research (TMLR)},
  year={2024},
  note={arXiv:2403.07815}
}

@article{das2024timesfm,
  title={A decoder-only foundation model for time-series forecasting},
  author={Das, Abhimanyu and Kong, Weihao and Sen, Rajat and Zhou, Yichen},
  journal={Proceedings of the 41st International Conference on Machine Learning (ICML)},
  year={2024},
  note={arXiv:2310.10688}
}

@article{woo2024moirai,
  title={Unified Training of Universal Time Series Forecasting Transformers},
  author={Woo, Gerald and Liu, Chenghao and Kumar, Akshat and Xiong, Caiming and Savarese, Silvio and Sahoo, Doyen},
  journal={Proceedings of the 41st International Conference on Machine Learning (ICML)},
  year={2024},
  note={arXiv:2402.02592}
}

@article{rasul2023lagllama,
  title={Lag-Llama: Towards Foundation Models for Probabilistic Time Series Forecasting},
  author={Rasul, Kashif and Ashok, Arjun and Williams, Andrew Robert and Ghonia, Hena and Bhagwatkar, Rishika and Khorasani, Arian and Bayazi, Mohammad Javad Darvishi and Adamopoulos, George and Riachi, Roland and Hassen, Nadhir and Bilo{\v{s}}, Marin and Garg, Sahil and Schneider, Anderson and Chapados, Nicolas and Drouin, Alexandre and Zantedeschi, Valentina and Nevmyvaka, Yuriy and Rish, Irina},
  journal={arXiv preprint},
  year={2023},
  note={arXiv:2310.08278}
}

@article{goswami2024moment,
  title={MOMENT: A Family of Open Time-series Foundation Models},
  author={Goswami, Mononito and Szafer, Konrad and Choudhry, Arjun and Cai, Yifu and Li, Shuo and Dubrawski, Artur},
  journal={Proceedings of the 41st International Conference on Machine Learning (ICML)},
  year={2024},
  note={arXiv:2402.03885}
}

@article{garza2023timegpt,
  title={TimeGPT-1},
  author={Garza, Azul and Challu, Cristian and Mergenthaler-Canseco, Max},
  journal={arXiv preprint},
  year={2023},
  note={arXiv:2310.03589}
}

@article{ekambaram2024ttm,
  title={Tiny Time Mixers (TTMs): Fast Pre-trained Models for Enhanced Zero/Few-Shot Forecasting of Multivariate Time Series},
  author={Ekambaram, Vijay and Jati, Arindam and Dayama, Pankaj and Mukherjee, Sumanta and Nguyen, Nam H. and Gifford, Wesley M. and Reddy, Chandra and Kalagnanam, Jayant},
  journal={Advances in Neural Information Processing Systems (NeurIPS)},
  year={2024},
  note={arXiv:2401.03955}
}

@article{aksu2024gifteval,
  title={GIFT-Eval: A Benchmark for General Time Series Forecasting Model Evaluation},
  author={Aksu, Taha and Woo, Gerald and Liu, Juncheng and Liu, Xu and Bian, Yuxuan and Liu, Chenghao and Savarese, Silvio and Xiong, Caiming and Sahoo, Doyen},
  journal={arXiv preprint arXiv:2410.10393},
  year={2024}
}

@article{assimakopoulos2000theta,
  title={The theta model: a decomposition approach to forecasting},
  author={Assimakopoulos, Vassilis and Nikolopoulos, Konstantinos},
  journal={International Journal of Forecasting},
  volume={16},
  number={4},
  pages={521--530},
  year={2000},
  publisher={Elsevier}
}

@book{hyndman2008ets,
  title={Forecasting with Exponential Smoothing: The State Space Approach},
  author={Hyndman, Rob J. and Koehler, Anne B. and Ord, J. Keith and Snyder, Ralph D.},
  year={2008},
  publisher={Springer}
}

@article{hyndman2006another,
  title={Another look at measures of forecast accuracy},
  author={Hyndman, Rob J. and Koehler, Anne B.},
  journal={International Journal of Forecasting},
  volume={22},
  number={4},
  pages={679--688},
  year={2006},
  publisher={Elsevier}
}

@article{wang2006characteristic,
  title={Characteristic-based clustering for time series data},
  author={Wang, Xiaozhe and Smith, Kate and Hyndman, Rob},
  journal={Data Mining and Knowledge Discovery},
  volume={13},
  number={3},
  pages={335--364},
  year={2006},
  publisher={Springer}
}

@article{cleveland1990stl,
  title={STL: A seasonal-trend decomposition procedure based on loess},
  author={Cleveland, Robert B. and Cleveland, William S. and McRae, Jean E. and Terpenning, Irma},
  journal={Journal of Official Statistics},
  volume={6},
  number={1},
  pages={3--73},
  year={1990}
}

@article{makridakis2018m4,
  title={The M4 Competition: Results, findings, conclusion and way forward},
  author={Makridakis, Spyros and Spiliotis, Evangelos and Assimakopoulos, Vassilios},
  journal={International Journal of Forecasting},
  volume={34},
  number={4},
  pages={802--808},
  year={2018},
  publisher={Elsevier}
}

@article{makridakis2020m4,
  title={The M4 Competition: 100,000 time series and 61 forecasting methods},
  author={Makridakis, Spyros and Spiliotis, Evangelos and Assimakopoulos, Vassilios},
  journal={International Journal of Forecasting},
  volume={36},
  number={1},
  pages={54--74},
  year={2020},
  publisher={Elsevier}
}

@article{makridakis2022m5,
  title={The M5 competition: Background, organization, and implementation},
  author={Makridakis, Spyros and Spiliotis, Evangelos and Assimakopoulos, Vassilios},
  journal={International Journal of Forecasting},
  volume={38},
  number={4},
  pages={1325--1336},
  year={2022},
  publisher={Elsevier}
}

\end{document}